\documentstyle[twocolumn,aps,epsfig]{revtex}

\begin{document}
\draft
\preprint{July 24, 2001}
\title{ Singular Density of States of 
Disordered Dirac Fermions in the Chiral Models}
\author{Shinsei Ryu$^{1}$ and Yasuhiro Hatsugai$^{1,2}$}
\address{
$^1$Department of Applied Physics, University of Tokyo,
7-3-1 Hongo Bunkyo-ku, Tokyo 113-8656, Japan\\
$^2$PRESTO, JST, Saitama 332-0012, Japan}

\maketitle

\date{\today}

\begin{abstract}
The Dirac fermion
in the random chiral models is studied
which includes
the random gauge field model and 
the random hopping model.
We focus on a connection 
between continuum and lattice models
to give a clear perspective
for the random chiral models.
Two distinct structures of density of states (DoS)
around zero energy,
one is a power-law dependence on energy 
in the intermediate energy range
and the other is a diverging one at zero energy,
are revealed by
an extensive numerical study
for large systems up to $250\times 250$.
For the random hopping model,
our finding of the diverging DoS
within very narrow energy range
reconciles previous inconsistencies
between the lattice and the continuum models.
\end{abstract}

\narrowtext

Dimensionalities and symmetries 
play central roles
for universalities
in the Anderson localization problem.
More than two decades ago,
Abrahams, Anderson, Licciardello and Ramakrishnan
\cite{abrahams}
presented
the well-known scaling theory,
predicting that 
electron wave functions always localize
in one and two dimensions
and metal-insulator transition occurs 
in three dimensions.
However, even in two dimensions,
delocalized states are also 
marginally allowed to appear
when systems possesses some special symmetries.
For symmetries, it is often convenient to borrow
terminologies from the random matrix theory 
which was first introduced by Wigner and Dyson.
Recently, Altland and Zirnbauer \cite{altland}
have reported seven new symmetry classes
in conjunction with 
mathematical classification scheme
of the Riemannian symmetric spaces.

Amongst these new symmetry classes,
chiral models have attracted much attention
as a novel exception for the scaling theory.
Compared to the conventional models where randomness
enters as an on-site potential,
it resides on {\it links } (i.e., as a gauge field) in chiral models.
This type of randomness may play important roles
for the composite fermion theory of fractional quantum Hall effects
 and vortex states of dirty superconductors.
In view of a localization problem,
models with this randomness have 
a special symmetry, referred as chiral symmetry,
and thereby belong to a new universality class \cite{altland}.
This symmetry is expected to affects 
localization properties of the systems drastically.

Several chiral models on a lattice, which are convenient 
for numerics, have been studied 
and interesting physics around zero-energy have been 
revealed.
DoS of these models show singularities at zero-energy
and the corresponding wave functions exhibit
a delocalized multifractal behavior.
Examples of these include
the Gade's model \cite{gade},
the random flux model \cite{sugiyama},
and the $\pi$-flux model with link disorders \cite{hatsugai}.
These models, defined on a two dimensional square lattice, 
have the chiral symmetry which
is conveniently stated
as
$\{ H, \gamma \}=0$
where $\gamma$ is a matrix 
which changes the sign of wave functions on a sub lattice.
Consequently, for any realization of the disorder, the energy spectrum
is invariant under the transformation $E\rightarrow -E$.
Therefore, given an eigenstate $\psi$ with energy $E$, 
$\gamma \psi$ is also 
an eigenstate with energy $-E$.
This symmetry is responsible for 
the existence of delocalized states 
at zero-energy.

Chiral models in a continuum space have also been investigated 
extensively.
Especially, 
models which include the Dirac fermions
have attracted an interest.
The Dirac fermion is a quasi particle which appears
in several interesting situations in condensed matter physics such as
$d$-wave superconductivity, graphite sheets, 
the gap-closing transition in quantum Hall effects\cite{hatsugai2}, 
the Chalker-Coddington network model\cite{ho},
and the mean field theory of the $t-J$ model\cite{affleck}.
Effects of randomness are also of fundamental interest
in these contexts.
Since many analytic approaches such as a field theoretic one are
applicable for the Dirac fermion,
several interesting results have been obtained so far.
Amongst them, there exist simple models which allow us to
construct an explicit zero-mode wave function for any realization of disorder
\cite{ludwig}.
Due to this advantage,
it has been revealed that
the zero-mode wave functions are not localized and
exhibit a multifractal behavior like ones in the lattice models.
Moreover,
there exists a {\it transition} in the multifractal spectrum
\cite{chamon}
and the density-density correlation \cite{ryu} as the disorder strength varied.

Although the delocalized multifractal nature of the exact zero-energy states 
has been well established now,
our knowledge for finite energy states, especially for
DoS around zero-energy,
is still in confusion.
For example,
a continuum model where the Dirac fermions feel random gauge fields
was studied in Ref. \cite{ludwig,mudry} and it was found that 
DoS exhibits a power-law dependence on energy.
On the other hand, a similar model with species doubling was 
studied in Ref.\cite{guruswamy} and diverging DoS was found.
From the lattice point of view, the corresponding model
including Dirac fermions
has not been studied so far.
However, the random flux model, 
a lattice model where flux is randomly distributed for each plaquette,
shows diverging DoS \cite{furusaki}.
A clear link between them is missing.

Another example is the case where the Dirac fermion
feels imaginary vector potentials.
For this case, there even exist inconsistent results between
lattice and continuum models.
Field theoretic studies \cite{guruswamy,fukui}
predicts diverging DoS at zero-energy for any randomness strength.
However, a numerical result for the random $\pi$-flux model,
where a random hopping amplitude act 
as a imaginary vector potential,
shows that DoS behaves as a power-law with its exponent dependent 
on the disorder strength for weak randomness \cite{hatsugai}.

In this paper, we make an attempt to 
clarify the relationship 
between lattice models and continuum Dirac fermions,
reconcile some inconsistent results in the previous studies,
and thereby give a whole perspective 
for the localization problem of the chiral models.
For these purposes,
we consider 
lattice models which recover the random Dirac fermions
in the continuum limit.
We focus on DoS, especially around zero energy
where quantum interference play
an important role
and fully quantum mechanical treatments are necessary.
We use the transfer-matrix method
developed in Ref.\cite{schweitzer}.
It allows us to handle
large enough systems up to $250\times 250$
, which is indispensable to give reliable results 
for localization problems.

We realize Dirac fermions on a 2D square lattice via the $\pi$-flux model:
\begin{eqnarray}
{\cal H}_{pure}=\sum_{<ij>}c_{i}^{\dagger}t^{pure}_{ij}c_{j}+h.c.
\nonumber
\end{eqnarray}
where
$
t^{pure}_{j+\hat{x},j}=(-)^{j_{x}}
$
,
$
t^{pure}_{j+\hat{y},j}=1
$
and flux piercing a plaquette is $\pi$.
The energy spectrum is given by
$
E=\pm 2 \sqrt{ \cos^{2}k_{x}a+ \cos^{2}k_{y}a }
$
where ${\bf k}$ belongs
to the magnetic Brillouin zone $(-\pi /a,\pi /a]\times(0,\pi/a ]$
and $a$ is a lattice constant.
In the continuum limit $a\rightarrow 0$, 
this includes the doubled massless Dirac fermions
around ${\bf k}=(\pi/2a,\pi/2a)$ and $(-\pi/2a,\pi/2a)$.
This realization 
is a minimum model for our purposes.

In the following, we will consider two types of disorder
which live on a link:
the random gauge field and the random hopping.
We implement the random gauge field as 
\begin{eqnarray}
\label{rf}
t^{rg}_{ij}=t^{pure}_{ij}\exp[i a A_{ij}].
\end{eqnarray}
Taking a Coulomb gauge, we determine 
the random gauge field $A_{ij}$ via a scalar potential $\Phi$
on a dual lattice
as
$
A_{j+\hat{x},j}= 
\left(  
\Phi_{j+(\hat{x}+\hat{y})/2}-\Phi_{j+(\hat{x}-\hat{y})/2} 
\right)/a
,
A_{j+\hat{y},j}=
-\left( 
\Phi_{j+(\hat{x}+\hat{y})/2}-\Phi_{j-(\hat{x}-\hat{y})/2} 
\right)/a
$
where $\Phi$ is randomly chosen from a simple Gaussian distribution
$ P [\Phi ] \propto 
\exp [-\frac{a^{2}}{2g} \sum_{<ij>}
\left( \Phi_{i}-\Phi_{j} \right)^{2}/a^{2}]$.
In the continuum limit, 
this reduces to $ P [\Phi ] \propto 
\exp [-\frac{1}{2g} \int d^{2}x
\left( \nabla \Phi \right)^{2} ]$,
a natural choice 
for effective field theoretic treatments.
As stated above, this Hamiltonian possesses the chiral symmetry
$\{ {\cal H}, \gamma \}=0$ due to the special nature of randomness.
Note also that time-reversal invariance is broken 
by the random magnetic flux.

In the continuum limit, the Hamiltonian is expressed as
\cite{hatsugai,guruswamy}
\begin{eqnarray}
\label{cont}
{\cal H}_{cont}= 
\int d^{2}x \Psi^{\dagger} 
\pmatrix{
0            & D    \cr
D^{\dagger}  & 0    \cr
} 
\Psi
\end{eqnarray}
where
$
D:= 
-i\partial_{x} \sigma_{x}
-i\partial_{y} \sigma_{y}
+A_{x}({\bf x})         \sigma_{x}
+A_{y}({\bf x})         \sigma_{y}
+M  ({\bf x})            \sigma_{z}
+V ({\bf x})            {\bf 1}
$ 
,
$\sigma_{i=x,y,z}$ are the $2\times 2$ Pauli matrices,
and $\Psi$ is a four-component spinor.
The coefficients 
$A_{x},A_{y},M$ and $V$ are arbitrary complex fields.
This is the most general chiral symmetric form 
of a Dirac Hamiltonian in a continuum space. 
The chiral symmetry is, in the present basis,
expressed as 
$\{ {\cal H},\sigma_{z}\otimes {\bf 1} \}=0$. 
For the random gauge field model,
$(A_{x},A_{y})$ is real and serves as the random gauge field.
The other coefficients  $M$ and $V$
do not appear at the first order in the lattice constant $a$,
but they are non-zero for the lattice models in general.

\begin{figure}
\epsfig{file=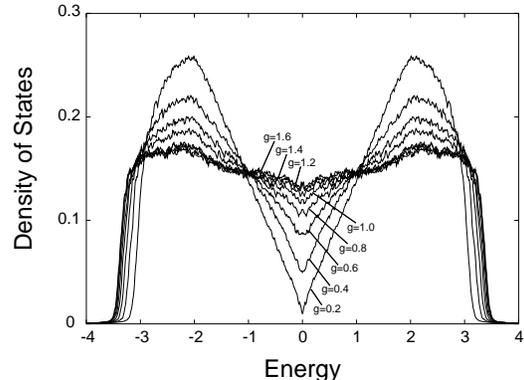,width=7cm}
\caption{
\label{rf1}
Density of states for the $\pi$-flux model 
with a random gauge field
on a $50 \times 50$ lattice
for $g=0.2-1.6$ (from bottom to top at $E=0$).
The small imaginary part of energy $\delta$ is 0.01.
Quenched averaging is taken over $30$ samples.
}
\end{figure}

\begin{figure}
\epsfig{file=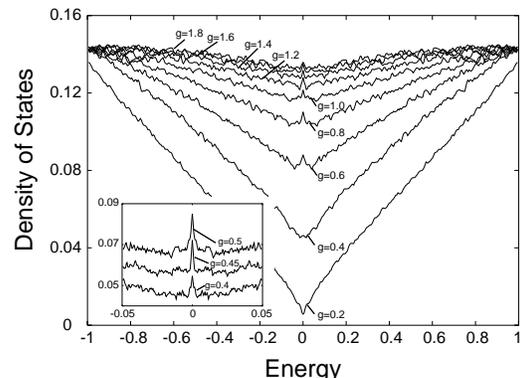,width=7cm}
\caption{
\label{rf2}
Same as Fig. \ref{rf1}
on a $100 \times 100$ lattice
for $g=0.1-1.8$, and $\delta=0.005$,
averaged over $40$ samples.
Inset:
same as Fig. \ref{rf1}
on a $250 \times 250$ lattice
for $g=0.4,0.45,0.5$, and $\delta=0.0005$,
averaged over $50$ samples.
}
\end{figure}

For strong enough randomness,
this model is naturally related to another lattice model, 
the random flux model.
(By the random flux model, we mean 
the model discussed in Ref.\cite{sugiyama,furusaki}
which does not include Dirac fermions,
while we simply call the model with Dirac fermions defined above
as the random gauge field model.)
To see this, note that the spatial correlation of the flux per plaquette
$\phi_{i}:=a^{2}\Delta_{i} \Phi_{i}$
is expressed as
$\left<\phi_{i}\phi_{j}\right> =
a^{4} \Delta_{i}\Delta_{j} G(|i-j|)$
where $G(|i-j|):=\left< \Phi_{i} \Phi_{j}\right>$ is
a lattice Green's function for $\Phi$,
and $\Delta_{i}$ is lattice Laplacian at a site $i$.
Taking a continuum limit $a\rightarrow 0$,
we find $\left<\phi_{i}\phi_{j} \right>=0$ for 
finite $a|i-j|$, i.e.,
the spatial correlation of flux is short ranged.
Moreover, for strong enough randomness,
lattice fermions do not remember original flux $\pi$ anymore,
and we expect this model to exhibit a similar behavior to that
of the random flux model where the flux for each plaquette is 
an independent random variable.

An effective sigma model via supersymmetry (SUSY) technique  
was studied in Ref. \cite{guruswamy},
predicting diverging DoS for all range of randomness strength.
A similar model without species doubling
$
{\cal H}=
\int d^{2}x \Psi^{\dagger} 
\left[ {\bf \sigma}\cdot{\bf p}+{\bf \sigma}\cdot {\bf A} \right]
\Psi
$
has also been studied \cite{ludwig,mudry} and power-law DoS 
with a disorder-dependent exponent was proposed.
On the other hand, the random flux model,
which is expected to be the strong randomness limit
of the present case,
shows a diverging DoS \cite{furusaki}.

\begin{figure}
\epsfig{file=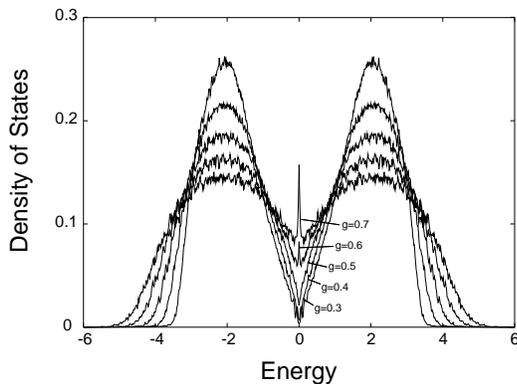,width=7cm}
\caption{
\label{rh1}
Density of states for the random hopping model
on a $50 \times 50$ lattice
for $g=0.3-0.7$ (from bottom to top at $E=0$).
The small imaginary part of energy $\delta$ is 0.01.
Quenched averaging is taken over $30$ samples.
}
\end{figure}

\begin{figure}
\epsfig{file=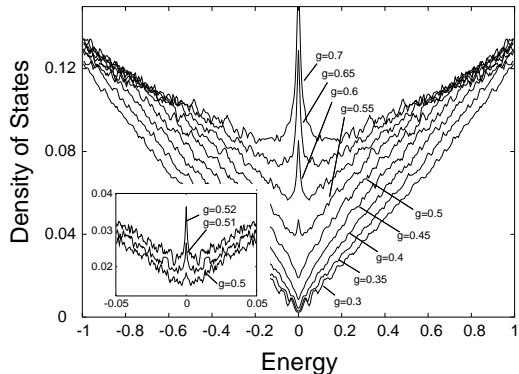,width=7cm}
\caption{
\label{rh2}
Same as Fig. \ref{rh1}
on a $100 \times 100$ lattice
for $g=0.3-0.7$, and $\delta=0.005$,
averaged over $50$ samples.
Inset:
same as Fig. \ref{rh1}
on a $250 \times 250$ lattice
for $g=0.5,0.51,0.52$,
and $\delta=0.0005$,
averaged over $70$ samples.
}
\end{figure}

We also consider the case 
where fermions
on a 2D square lattice with $\pi$-flux per plaquette
feel random hopping amplitudes:
\begin{eqnarray}
\label{rh}
t^{rh}_{ij}=t^{pure}_{ij}+\delta t_{ij}
\end{eqnarray}
where $\delta t_{ij}$ is a real random variable.
As for a probability distribution of $\delta t$,
we consider a Gaussian distribution
$
P[\delta t] \propto \exp [ -(\delta t)^{2}/ 2 g^{2} ]
$.

In this case, 
$(A_{x},A_{y})$ 
in Eq.(\ref{cont}) is purely imaginary,
and Dirac fermions feel an imaginary random gauge potential.
Note that the total four-component Hamiltonian is 
hermitian.
Unlike the random gauge field model,
time-reversal symmetry is not broken
for this model
since we can take all matrix elements to be real.

As is the case of the random gauge field model,
this model is connected to another lattice model
for sufficiently strong randomness.
For strong enough randomness, again, one expects
that fermions do not remember the original flux
anymore and shows a similar behavior
to that of the Gade's model.

Fukui \cite{fukui} studied the continuum model 
by a replica nonlinear sigma model
with a large number of fermion flavors
and proposed that DoS at zero-energy diverges
and delocalized states exist at the band center
for any randomness strength.
One-loop renormalization group (RG) study 
with SUSY method
was also applied
for this model in Ref. \cite{guruswamy},
predicting diverging DoS at zero energy as
$
\rho(E)
\sim \frac{E_{R}}{E}
e^{
-c \sqrt{\ln (E_{R}/E)}
}
$
where $E_{R}$ is a constant and $c$ depends on randomness.
A conjectural RG flow beyond one-loop order
was presented 
in Ref. \cite{leclair}
and similar results were obtained.
The Gade's model, which we expect to be the strong randomness
limit of the present model,
also shows diverging DoS \cite{gade}.
On the other hand, according to a numerical study 
for the corresponding lattice model,
vanishing DoS $\rho(E)\sim |E|^{\alpha}$ was observed
with a disorder-dependent exponent $\alpha >0$
for {\it weak} randomness \cite{hatsugai}.
We re-examined this problem by the transfer-matrix method.

A quantity of interest we investigate numerically is random averaged DoS:
$
\left< \rho(E) \right>:=
\left< \frac{1}{L^{2}}\sum_{i}\delta(E_{i}-E) \right>
$.
For the pure $\pi$-flux model, DoS vanishes linearly around zero-energy
due to the relativistic dispersion of Dirac fermions.
Since for the strong enough disorder,
the present models are expected to exhibit similar properties
to those of the random flux model or the Gade's model,
both of which exhibit divergent DoS at zero-energy\cite{gade,furusaki},
a natural question is how the vanishing DoS at zero-energy 
becomes divergent as we increase the disorder strength $g$.
It should be contrasted to the usual case
where DoS of pure systems are also divergent 
at zero-energy due to the van-Hove singularities
of the tight binding model on a two-dimensional square lattice.

First, we discuss the random gauge field model,
i.e., $\pi$-flux model with a random gauge field as in Eq. (\ref{rf}).
In Fig. \ref{rf1} and \ref{rf2},
we present numerically calculated DoS for several $g$.
In the transfer matrix method,
the Green's function for a given energy $E$ is calculated
at $E+i\delta$.
We chose $\delta \sim 0.01-0.0005$ which
gives us enough resolution of energy.
We used a square geometry rather than a quasi one-dimensional one
since we are mainly interested in pure two-dimensional properties.

As shown, there exist two structures.
For over the wide range of the energy scale,
the global feature of DoS
changes from V-shaped to flat one,
the latter
is characteristic to the random flux model
as expected.
In addition, if we look more precisely,
for very tiny region around zero-energy,
there exists another structure,
i.e., 
diverging DoS for sufficiently large $g$.
Note that 
since we used a lattice with even number of sites 
and adopted periodic boundary condition,
there is no exact zero energy state,
i.e., the diverging DoS found here 
is not an artifact by the special choice
of boundary conditions.
The numerical results found here 
are 
consistent with 
the sigma model study via SUSY technique  
in Ref. \cite{guruswamy},
where the divergence of DoS is predicted.

Now let us go on to the $\pi$-flux model with 
random hopping amplitudes ( Eq.(\ref{rh}) ).
Calculated DoS for systems up to $250\times 250$
are shown in Fig.\ref{rh1} and \ref{rh2}.
Again, one recognizes two different structures as 
in the random gauge field model.
Away from zero energy,
DoS exhibits a power law behavior
with a disorder-dependent exponent
as in Ref. \cite{hatsugai}.
The whole DoS profile becomes divergent as the disorder strength increased,
which is natural since the present model is 
expected to show similar behaviors to that of the Gade's 
model.
On the other hand, 
as is the case with the random gauge field model studied above,
within a very narrow range near zero-energy,
there exists another structure.
As the disorder strength increased,
DoS divergence at zero-energy occurs  
when $g$ reaches about the order of the bandwidth.
This behavior is independent of 
the global power-law profile away from the zero energy
and before it turns to be divergent.
These findings are consistent with the field theoretic analyses, 
especially with the SUSY approach 
where the divergence of DoS is predicted
and a power-law behavior appears in the intermediate region
of the RG flow.
We also investigated the case where
$\delta t$ is uniformly distributed between $[-w/2,w/2]$.
The results for this case are qualitatively similar to those of a Gaussian 
distribution.

We should compare the present case to dirty $d$-wave superconductors
which also include Dirac fermions,
however, the underlying symmetry class is different 
in the random matrix theory
\cite{altland}.
For dirty $d$-wave superconductors,
all quasiparticle states are found to be localized,
and 
a small energy scale around zero energy 
appearers
where quantum interference effects produce
several critical DoS profiles depending on 
the details of randomness \cite{atkinson}.
This energy scale is determined by
the diffusion constant in the diffusive regime
and 
the localization length of quasiparticles \cite{senthil}.
For the present case, on the other hand, 
zero-energy states are delocalized due to 
the chiral symmetry,
and quantum interference effects
give rise to diverging DoS at zero energy.
The emergence of 
a power-law behavior away from zero energy is, 
from the effective field theory point of view,
well described by
the intermediate regime of the RG flow
\cite{guruswamy}.

In conclusion, 
we have investigated DoS of the chiral models
where the chiral symmetry plays crucial roles
for the existent of the delocalized states at
zero energy.
We considered the lattice counter parts
of continuum models including
the Dirac fermions,
the random gauge field model and
the random hopping model,
which
have been well studied by field theoretic
methods.
Large-scale calculations by the transfer matrix method 
up to $250 \times 250$ lattices
revealed 
the existence of two distinct structures in DoS,
one is a power-law behavior in the intermediate energy range
and the other is a diverging one around zero energy.
Quantum interference plays crucial roles for these
fine structures, 
which naive semiclassical treatments may miss.
Our finding of the diverging DoS at zero-energy
for the random hopping model
reconciles inconsistencies between 
lattice and continuum models
and thus establishes a clear connection
between them.

We thank Y. Morita for fruitful discussions.
The computation in this work has been partly done 
at the YITP Computing Facility 
and at the Supercomputing Center, ISSP, University of Tokyo.

\end{document}